\documentclass[12pt]{article}

\usepackage{scicite}

\usepackage{times}
\usepackage{graphicx}
\usepackage{url}
\usepackage{booktabs}
\usepackage[all]{nowidow}

\newenvironment{sciabstract}{%
\begin{quote} \bf}
{\end{quote}}

\title{ Social Approval and Network Homophily as Motivators of Online Toxicity}

\author
{Julie Jiang,$^{1\ast}$ Luca Luceri,$^{1}$ Joseph B Walther,$^{2}$ Emilio Ferrara$^1$\\
\\
\normalsize{$^{1}$University of Southern California}\\
\normalsize{$^{2}$University of California, Santa Barbara}\\
\\
\normalsize{$^\ast$juliej@isi.edu}
}

\date{}
\begin{document} 

% Double-space the manuscript.

\baselineskip24pt

% Make the title.

\maketitle

% Place your abstract within the special {sciabstract} environment.

\begin{sciabstract}
Online hate messaging is a pervasive issue plaguing the well-being of social media users. This research empirically investigates a novel theory positing that online hate may be driven primarily by the pursuit of social approval rather than a direct desire to harm the targets. Results show that toxicity is homophilous in users’ social networks and that a user’s propensity for hostility can be predicted by their social networks. We also illustrate how receiving greater or fewer social engagements in the form of likes, retweets, quotes, and replies affects a user’s subsequent toxicity. We establish a clear connection between receiving social approval signals and increases in subsequent toxicity. Being retweeted plays a particularly prominent role in escalating toxicity. Results also show that not receiving expected levels of social approval leads to decreased toxicity. We discuss the important implications of our research and opportunities to combat online hate.
\end{sciabstract}

%\keywords{Keyword1, Keyword2, Keyword3}

\section*{Introduction}

The proliferation of hate messages in social media–commonly understood as expressions of hatred, discrimination, or attacks towards individuals or groups based on identity attributes such as race, gender, sex, religion, ethnicity, citizenship, or nationality \cite{tsesis2002destructive}--has garnered considerable research attention over recent years \cite{paz2020hate,ezeibe2021hate,thomas2021sok,frimer2023incivility}. Online hate has important connections to cyberbullying \cite{chen2011detecting} and online harassment \cite{thomas2021sok} and, in extreme cases, to the incitement of violence and offline hate crimes \cite{ezeibe2021hate,castano2021internet,muller2021fanning,wang2023identifying}. Research finds that young adults, LGBTQ+ minorities, and active social media users are especially vulnerable to online hate and harassment \cite{keipi2016online,thomas2021sok}. Moreover, evidence suggests the pervasiveness of online hate is growing increasingly \cite{mathew2020hate,thomas2021sok,frimer2023incivility}, which emphasizes the urgency to address problems of online hate and toxicity.
% \footnote{In this paper, we use the terms ``online hate'' and ``toxicity'' interchangeably.} 

This research provides an initial empirical test of a new theory focusing on the motivations and gratifications associated with posting hate messages online. It posits that online hate is fueled by the social approval that hate message producers receive from others \cite{walther2022social}. The theory suggests that online hate behavior is not primarily motivated by the desire to harm prospective victims but rather to accrue validation and encouragement from like-minded others. Within this framework, one’s propensity to express hateful messages should be related to a similar propensity among one’s social network; people who share similar resentments and actions should be linked, which can facilitate mutual reinforcement of one other’s hate behavior. Further, their hate messaging would be expected to become more extreme as they obtain more reinforcement through social approval signals from others, potentially in the form of likes, upvotes, or other forms of positive feedback. 

The contribution of this work is two-fold. First, we investigate whether online hateful behavior conforms to patterns of social network homophily--the idea that people who share similar characteristics, interests, or behaviors are also frequently associated with one another--is a phenomenon repeatedly observed in many settings \cite{mcpherson2001birds,kossinets2009origins}. Some suggest that the expression of hate is also a homophilous trait \cite{nagar2022homophily}. In a comparison study of hateful and non-hateful Twitter users, prior work has shown that hateful users have higher network centrality \cite{ribeiro2017like}. Additionally, semantic, syntactic, stylometric, and topical similarities exist among hateful users connected in a Twitter network. Another work on COVID discourse on YouTube highlights that YouTube commentators are segregated by toxicity levels \cite{obadimu2021developing}. Relatedly, a study on moral homophily found that moral convergence in a cluster of users in an extremist social network predicts how often they spread hateful messages \cite{atari2022morally}. However, these works do not elucidate the extent to which social network cues facilitate detecting toxic users. Drawing upon notions of social network homophily, we explore whether we can utilize them to predict user toxicity labels from a smaller, more limited training set, which would suggest that homophily plays an important role in easily locating toxic users.

Our second contribution considers how the social network influences toxicity. Specifically, does receiving social approvals make a user who is already hateful more hateful? While several empirical studies report that tweets that are more uncivil, toxic, or otherwise outrageous generate more signals of social approval in the forms of likes and retweets  \cite{brady2017emotion,kim2021distorting,frimer2023incivility}, scant research has examined the opposite: whether the reception of social approval to one’s toxic messages encourages yet greater incivility and toxicity. This is the core proposition of the social approval theory of online hate: The audience for an author’s hate messages (that appear nominally focused on some target minority) is like-minded online peers and friends, whose signals of approval reinforce and encourage more extreme hatred in an author’s subsequent messages. One possible illumination of this dynamic is the work of Frimer et al. \cite{frimer2023incivility} examining the increased incivility among US politicians on Twitter. In particular, they found that politicians responded to likes and retweets for their uncivil tweets by escalating their toxicity in the future \cite{frimer2023incivility,brady2021social,shmargad2022social}. However, whether these findings generalize to normal users is unclear since politicians may be more pressured to respond to constituents’ approval than normal users. The present work helps to bridge this gap by examining the effects of social feedback on a more representative sample of ``average'' yet hateful Twitter users.

Using an extensive set of historical tweets by socially connected users who are known to be hateful, we conduct a large-scale analysis of the hateful messaging behavior. We present evidence of homophilous toxicity among hate-producing social networks of social media users. This corroborates previous work, which showed that social network cues of retweets and mentions (e.g., by quoting someone) can be leveraged to predict how hateful the users are \cite{jiang2023social}. In addition, we show that social approvals are linked to increased toxicity, whereas insufficient approvals are linked to decreased toxicity. Particularly, being retweeted--a form of social approval that signifies both endorsement and flattery--has the strongest link with increased hateful behavior. Receiving likes has a similar but smaller effect. Receiving \textit{fewer} replies or quotes–which, unlike retweets and likes, could contain criticism instead of approval--also leads to greater subsequent toxicity. This research carries important theoretical implications to advance our understanding of how social gratifications affect the propagation of online hate and suggest alternative strategies to deter it.  

\section*{Results}
In this work, we consider a large dataset of known, hateful users. We collect a relatively comprehensive collection of their original tweets, along with timestamps and social engagement counts. We apply a hate score detection model to detect the toxicity of each tweet and also average the score per user. Additionally, we build a social network graph consisting of pairs of users connected by retweets and/or mentions. Since many of the users in our dataset may be inauthentic users, we conduct most of our studies on users with bot scores $<=0.5$ (very likely to be human) and replicate them on users with bot scores $<=0.8$ (likely to be human). Please see the Methods section for more details below. 
\begin{table*}[t]
    \centering
    % \footnotesize
    
    \caption{Hate scores among users exhibit homophily in the social network, as indicated by both the network assortativity and the Pearson correlation between a user's hate score and the weighted average of their neighbors' (***$p<0.001$) .}
    \label{tab:toxicity_homophily}
    \begin{tabular}{ccrrrr}
    \toprule
    \textbf{Network} &  \textbf{Bot Score}  & \textbf{\# Nodes} & \textbf{\# Edges} & \textbf{Assort.} & \textbf{Wgt. Avg. Corr.}\\
    \midrule
      Retweet & $<=0.8$ &  6,665 & 74,943 &  0.071*** &  0.310*** \\
      Retweet & $<=0.5$ &  2,985 &   9,016 &  0.195*** &  0.445*** \\
      Mention & $<=0.8$ &  6,665 & 104,802 &  0.057*** &  0.244*** \\
      Mention & $<=0.5$ &  2,985 &  14,182 &  0.185*** &  0.421*** \\
    \bottomrule
    \end{tabular}
\end{table*}
\subsection*{Homophily in Toxic Behavior}\label{sec:rq1}
To investigate whether toxicity is a homophilous behavior, we compute network assortativity and weighted average correlation metrics based on the social network and the user hate scores. Both metrics use the Pearson correlation method. The results shown in Table \ref{tab:toxicity_homophily} demonstrate that the hate scores exhibit homophily in both the retweet and mention networks. We isolate users by two bot score thresholds, where the higher the bot score indicates the more likely the user is a bot, to eliminate the impact of bots. Both the network assortativity and the weighted average correlation of a user’s hate score and their neighbors are significantly positive.
% In particular, the weighted average correlation, which includes a weighted average calculation of a user's neighbors, is much greater, indicating the strength of connectivity among all neighbors is more correlated with a user's own toxicity. 
To ensure the validity of these findings, we perform a robustness check by comparing our results with those obtained from a null model. In the null model, we randomly shuffle the nodes on one end of the edges so that each actual edge $A\leftarrow B$ is now $A\leftarrow C$, where $C$ is a random node. This procedure yields nonsignificant Pearson correlations, all with absolute values less than 0.02 in every case, suggesting that the social network reflects homophily in toxicity. Interestingly, homophily is more evident when we consider only users with lower bot scores. One possible reason is that users who are more likely to be genuine users rather than social bots tend to create more homophilous connections (see the Supplementary Information for further analysis).

% The presence of homophily can be further corroborated by the work of \cite{jiang2023social}, which proposed a general-purpose user detection modeling framework \textit{Social-LLM} based on social network data. They validated their model on the same dataset used in this paper for user toxicity detection. They found that using the social network (e.g., who retweets who) combined with other simple user metadata features such as profile descriptions can predict users' hate scores much more accurately than other methods that do not utilize the network features. It additionally provides a set of user representation embeddings that can be used for any downstream tasks, which we will use in the following section to study the effect of social engagement on toxicity.

\begin{figure}[t]
    \centering
    \includegraphics[width=0.5\linewidth]{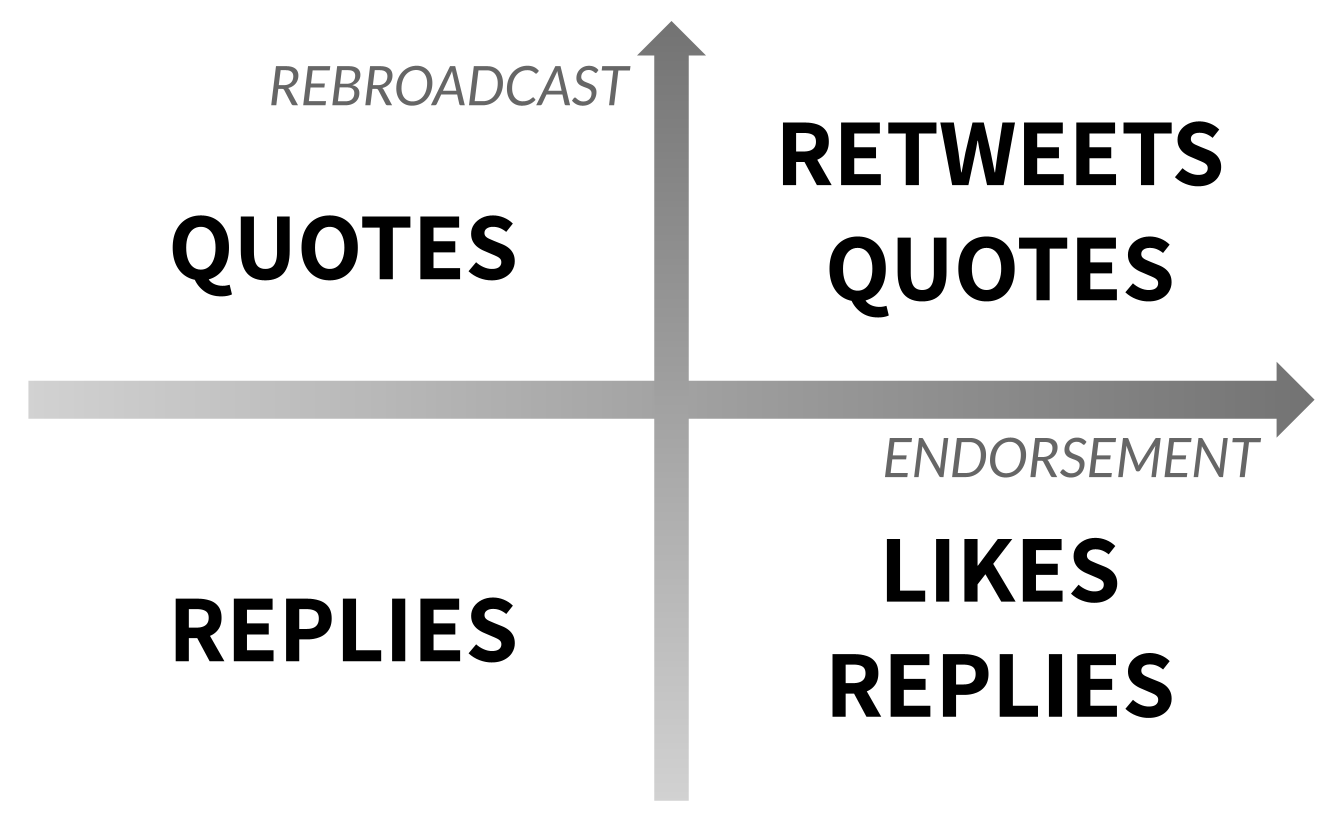}
    \caption{The four types of social engagement on dimensions of rebroadcast and endorsement. \textit{Retweets} represent rebroadcast and endorsement,  \textit{likes} represent endorsements, \textit{quotes} are rebroadcasts that can be either positive or negative, and \textit{replies} do not rebroadcast and can be either positive or negative.}
    \label{fig:social_engagement_diagram}
\end{figure}
 
\subsection*{The Effect of Social Engagement on Toxicity} \label{sec:rq2}
Next, we analyze how social engagement may affect a user's propensities toward toxicity. Before presenting the results, let us review the four types of social engagement signals and what they potentially implicate. On Twitter, one user may engage with another user's post by liking, retweeting, quoting (retweeting with additional comments), or replying (commenting). As illustrated in Fig. \ref{fig:social_engagement_diagram}, we can position each form of social engagement jointly on the dimensions of rebroadcast and endorsement. By retweeting, one rebroadcasts a tweet to one’s own followers on Twitter. Retweeting may connote exceptional social approval, as a retweet not only signifies endorsement but the flattering reflection of another user’s desire for the original message to be seen by their own friends and followers \cite{boyd2010tweet,metaxas2015retweets}. Liking a tweet also signifies endorsement but not the flattery implied by re-transmission. Quotes are similar to retweets, except that a quoter adds additional comments of their own that can either support or disparage the original tweet \cite{hemsley2018tweeting}. Similarly, replies can support or undermine the original tweet, but replies do not rebroadcast the original tweet. Considering these possibilities, we contend that retweeting conveys the strongest degree of social approval, followed by likes. Quotes and replies can express social approval or disapproval, but without the textual content, it is impossible to determine which sentiment it conveys. Of course, an individual can interact with a tweet in multiple ways, for instance, by retweeting \textit{and} liking, but we cannot determine this from our data. In the absence of engagement metrics that explicitly signal disapproval (such as downvoting), we attempt to distinguish social approval from disapproval by how these metrics relate to one another. For example, a tweet with relatively more replies than likes could indicate that the tweet is perceived negatively by others. Conversely, if a tweet has both a high number of quotes and retweets, the tweet may be viewed favorably.

\begin{table*}[t]
    \centering
        
    \caption{Number of anchor tweets and their corresponding number of unique users (bot score $<=0.5$) when the engagement metric that the anchor tweet received is substantially lower or higher than predicted ($k=50$). }
    \label{tab:n_instances_rq2}
        \begin{tabular}{@{}lrrrr@{}}
        \toprule
        & \multicolumn{2}{c}{\textbf{Lower Than Predicted}} & \multicolumn{2}{c}{\textbf{Higher Than Predicted}} \\
        \cmidrule(lr){2-3} \cmidrule(lr){4-5}
        \textbf{Metric} & \textbf{\# Ex} &  \textbf{\# Users} & \textbf{\# Ex} &  \textbf{\#  Users} \\
        \midrule
        Likes & 8,962 & 741 & 15,835 & 1,447 \\
        Retweets & 16,969 & 1,062 & 63,779 & 2,022 \\
        Replies & 16,496 & 642 & 30,743 & 1,800 \\
        Quotes & 2,462 & 347 & 52,279 & 1,883 \\
        \bottomrule
        \end{tabular}
\end{table*}
To examine the potential impact of social engagement on toxicity, we focus on instances where users experienced social approvals or disapprovals that significantly deviated from their expectations. To achieve this, we construct four machine-learning models to predict four distinct social engagement signals per tweet (see Materials and Methods). For example, in estimating the number of likes a tweet garners, we incorporate metrics such as retweets, quotes, replies, along with an extensive set of other features derived from both user information and text content, aiming to predict the tweet's social engagement. This model can help identify tweets where the predicted engagement value substantially differs from the actual value, suggesting situations where users received unexpected levels of social approval or disapproval. We refer to these tweets as "anchor" tweets and analyze the average change in toxicity between the $k$ tweets preceding and following the anchor tweet. The following results presented focus on users with bot scores less than 0.5. For robustness, we replicate our findings for users with bot scores less than 0.8 in the Supplementary Information to demonstrate the consistency of our results. We display the number of anchor tweets and their corresponding unique users that received significantly higher or lower engagement counts for $k=50$ in Table \ref{tab:n_instances_rq2}.

\begin{figure}
    \centering
    \includegraphics[width=0.7\linewidth]{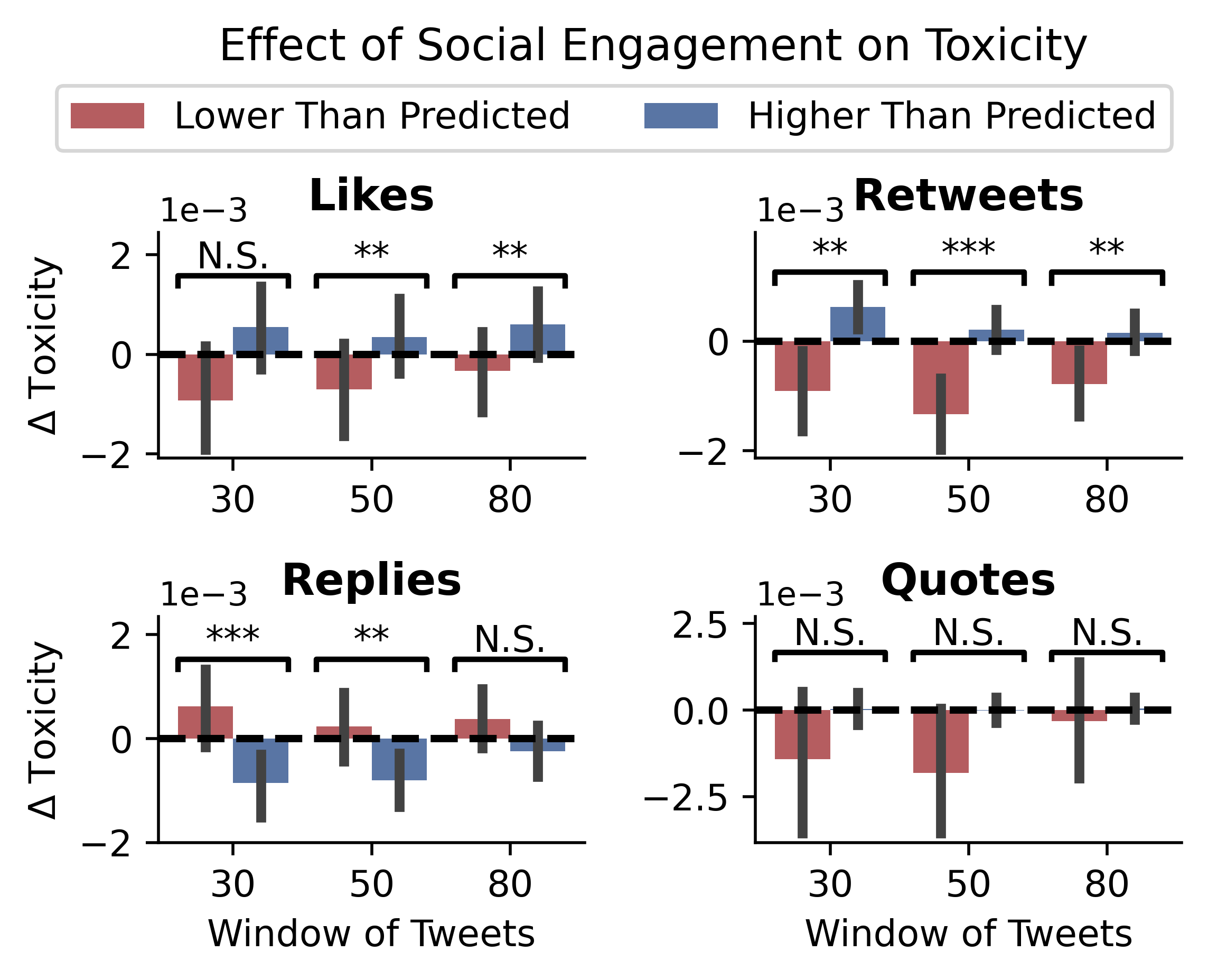}
    \caption{Changes in toxicity (y-axis) when an anchor tweet received lower (red bars) or higher (blue bars) than the predicted amount of social engagement at different windows $k$ (x-axis). Changes that are significantly different between the lower- and the higher-than-predicted groups are indicated (Mann-Whitney U test, ** $p<0.01$, *** $p<0.001$).}
    \label{fig:social_engagement_toxicity_diff}
\end{figure}

\subsubsection*{Likes and Retweets Increase Toxicity, but Replies Reduce It}
Fig. \ref{fig:social_engagement_toxicity_diff} illustrates the changes in a user’s toxicity when an anchor tweet experienced a substantially higher or lower amount of social engagement. Let $k=50$ indicate that we compare the past 50 tweets before the anchor tweet with the 50 tweets following the anchor tweet, we see that anchor tweets that received substantially more likes, more retweets, or fewer replies than expected lead to a \textit{significantly greater increase in toxicity} than when the anchor tweet received substantially fewer likes, fewer retweets, or more replies than expected. Though the net effect is relatively small, we emphasize the statistical significance of our results. In particular, not receiving enough retweets seems to have a much more dramatic effect on users: users who received substantially fewer retweets than expected generated a larger net decrease in subsequent toxicity (red bars) than if they received more retweets than expected (blue bars).

The effects of social engagement on toxicity may also reflect different temporal durations. While the increased toxicity due to retweets replicates for all other values of $k$, it differs for likes and replies. For likes, we observe a sustained increase in toxicity when the window is larger $k=50, 80$ but not smaller; for replies, the reduction of toxicity persists for smaller $k= 30, 50$ but not larger windows. One potential explanation is that giving likes requires little effort, and therefore ``likes'' connote less potent social approval compared to more effortful verbalized approval messages in the form of quotes or replies, so the effect of likes is more gradual and long-term. If a reply is negative, the explicit criticism can immediately affect a user’s behavior, but the effect may not be long-lasting.

\begin{table*}[t]
    \centering
    
    \caption{The average change in toxicity at $k=50$ when an anchor tweet received substantially higher or lower likes-per-quotes or retweets-per-quotes than expected. Statistical significance from a Mann-Whitney U test is indicated (** $p<0.01$, *** $p<0.001$).}
    \label{tab:relative_social_engagement_toxicity}
    \begin{tabular}{lrrc}
        \toprule 
         & \multicolumn{2}{c}{\textbf{Actual vs. Expected}} \\
         \cmidrule(lr){2-3}
        \textbf{Relative Metric} & \textbf{Lower} & \textbf{Higher} &  \textbf{Sig.}\\
        \midrule
         % Likes-per-Retweets & 0.0014 & 0.0001 & N.S.\\ 
         % Likes-per-Replies & $-0.0001$ &0.0009 & N.S.\\
         % \rowcolor{ACMYellow} 
         Likes-per-Quotes & $-0.0007$& 0.0005	& \textasteriskcentered\textasteriskcentered\\
         % Retweets-per-Replies &$-0.0058$& 0.0003& N.S. \\
         % \rowcolor{ACMYellow} 
         Retweets-per-Quotes  & $-0.0019$ &	0.0005	& \textasteriskcentered\textasteriskcentered\textasteriskcentered\\
         % Replies-per-Quotes & $-0.0005$& $-0.0009$& N.S. \\ 
         \bottomrule
    \end{tabular}

\end{table*}

\subsubsection*{Relatively Fewer Quotes Increase Toxicity}
It appears that quotes lead to no discernable changes in toxicity (Fig. \ref{fig:social_engagement_toxicity_diff}). However, it could be the case that they \textit{do} impact toxic behavior, yet because their impact could be both extremely negative or extremely positive, the net effect may be canceled out. To test this possibility, we look at the relative number of quotes compared to the number of likes or retweets, two engagement methods that we are quite certain to be approvals and not disapprovals. We train a new social engagement model to predict $x$-per-quote, where $x$ could be likes, retweets, or replies. We observe that when there are more likes-per-quotes or retweets-per-quotes, a user displays a significantly greater increase in toxicity (Table \ref{tab:relative_social_engagement_toxicity}). The effect is not significant when we compare replies-per-quotes. One possibility is that since likes and retweets are relatively unambiguous positive social approval, a high number of likes/retweets in conjunction with a low number of quotes indicates more social approval than disapproval, and vice versa.  

\begin{figure}
    \centering
    \includegraphics[width=0.5\linewidth]{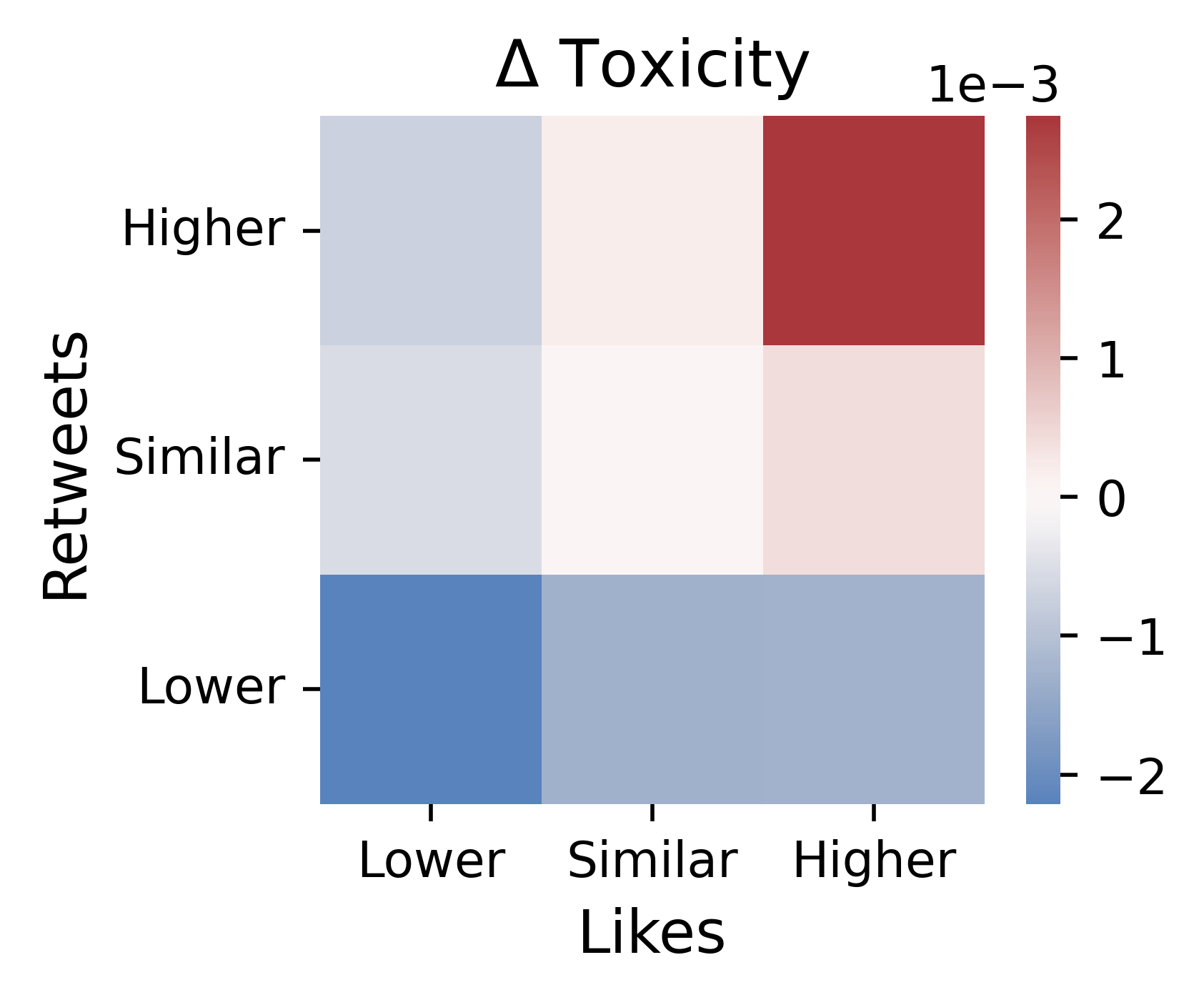}
    \caption{Having higher amounts of likes \textit{and} retweets than predicted would result in the biggest increase in future toxicity, and vice versa ($k=50$).}
    \label{fig:like_and_retweet_toxicity}
\end{figure}

\subsubsection*{The Compounded Impact of Both Likes and Retweets}
Thus far, we have only considered engagement metrics in isolation. However, a user would presumably be impacted by \textit{all} forms of social engagement on their post at once. Therefore, we analyze the combined effects of likes and retweets (Fig. \ref{fig:like_and_retweet_toxicity}). When both likes and retweets are greater than expected, the increase in subsequent toxicity doubles compared to when only one form is greater. The opposite is also true: fewer likes in conjunction with fewer retweets reduce toxicity by a greater amount. 

\begin{figure}
    \centering
    \includegraphics[width=0.5\linewidth]{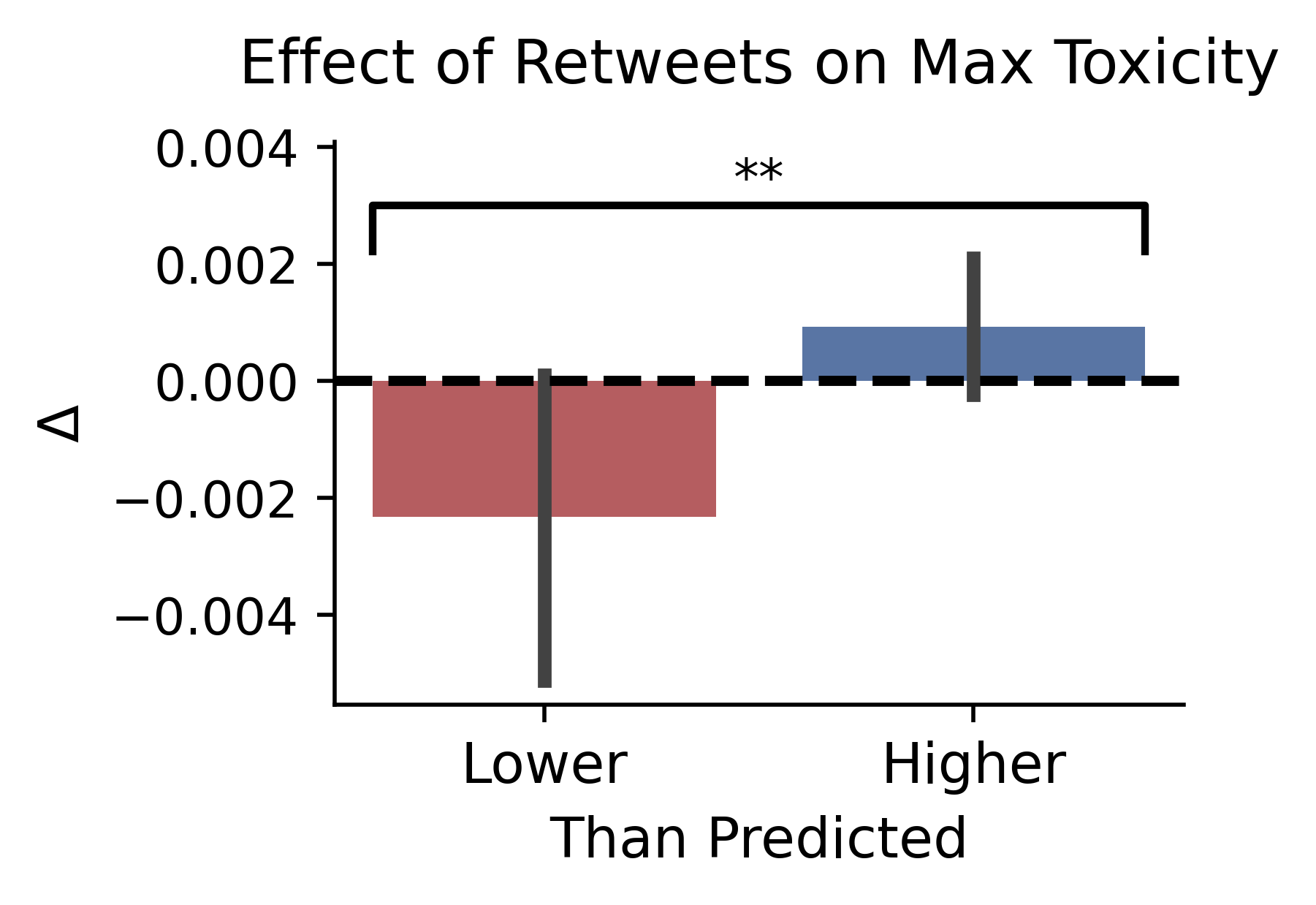}
    \caption{When an anchor tweet receives substantially lower (red) or higher (blue) amount of retweets than expected, the difference in maximum toxicity ($k=50$) is statistically significant (Mann-Whitney U test, $**p<0.01$). More retweets lead to an increase in maximum toxicity, and vice versa}
    \label{fig:retweet_max_toxicity}
\end{figure}
\subsubsection*{Retweets Escalate Maximum Toxicity} 

While we have shown that the toxicity levels are raised by social approval, we wonder whether social approval influences how hateful one is in one's most hateful tweet. Similar to our previous approach, we compare the maximum toxicity in the tweets before and after an anchor tweet. Fig \ref{fig:retweet_max_toxicity} shows that the change in maximum toxicity is statistically significant when the anchor tweet experiences substantially fewer or greater retweets. After an anchor tweet gets more retweets than expected, the user escalates their maximum toxicity. In contrast, after anchor tweets receive fewer retweets than expected, users decrease their maximum toxicity. This trend is only observed with retweets and not with likes, replies, or quotes (which are not significantly different, not shown), suggesting that retweets hold unparalleled power in altering online hate.

\section*{Discussion}

In this paper, we take a comprehensive look at millions of historical tweets by a set of known, hateful users on Twitter. We make two meaningful contributions. First, we show that toxicity is homophilous on social networks. Second, we find that a hateful user’s toxicity level can rise or fall when the user experiences substantially less or more social engagement in the form of \textit{retweets} (flattering rebroadcast and endorsement), \textit{likes} (endorsement), \textit{replies} (either endorsement or criticism), and \textit{quotes} (rebroadcast and either endorsement or criticism). We find that users’ tweets grow significantly more toxic after they receive more social approval from other users: more retweets, more likes, fewer replies, and comparatively fewer quotes. Conversely, users' tweets become less toxic after they receive insufficient social approval or social engagement that may indicate disapproval rather than approval. In particular, retweets, which signal both rebroadcast and endorsement, consistently is linked to a profound increase in users’ toxic behavior. These findings can be extended by analyzing the relation between social engagement signals and higher-level behavioral cues to explain the incentives and motivations of hateful actors through modeling techniques such as inverse reinforcement learning (e.g., \cite{luceri2020detecting}).

Our results support the social approval theory of online hate \cite{walther2022social}: hateful users could be motivated to appeal to their supporters and respond suitably to positive social reinforcement. It may be hateful users are \textit{not} primarily incentivized to harm the nominal target of their disparaging messages but rather to gain favor from their hateful peers. In addition to advancing our understanding of toxic online behavior, we believe this work has tremendous potential to inform strategies to combat hate speech on not only Twitter but also other online platforms. For one, this empirical evidence supports efforts to moderate hateful content (e.g., Meta’s policy \cite{fbpolicy}). Other directions include ``shadowbanning'' (hiding users’ posts from all but the user; see \cite{jaidka2023silenced}) or disabling likes, etc., on hateful posts to reduce the effects of social reinforcement on toxicity.

% \paragraph{Limitations}
Our research has limitations. Most importantly, while our findings are wholly consistent with the social approval theory, we recognize that we did not conduct any randomized, controlled experiment to demonstrate causality. Conducting such an experiment on this topic would not be ethical, as it would require us to reward hateful users for discriminatory messages and thereby encourage online hatred. Associational research with temporal order provides the best estimation of potentially causal theoretical relationships under the circumstances \cite{davis1985logic}. We also note other limitations due to data availability. Our data is based on one hate-infused dataset of Twitter posts concerning specifically US/UK immigration \cite{bianchi2022njh}, and our sample may be skewed since many users from the seed dataset are likely to be suspended by Twitter or have deleted their own accounts. Additionally, in the absence of the quote and reply texts that may either endorse or criticize a tweet, we cannot examine the full impact of social approval or disapproval signals on hateful behavior. Lastly, our findings regarding the impact of social engagement on future toxicity reveal small effect sizes. However, despite these shortcomings, the statistical significance and the consistency of our findings across various robustness tests enhance the validity of our conclusions. Many other influences impinge on the tenor of hate messages, from real-world geopolitical conflicts to mainstream media stories, not to mention changes in online content moderation policies. Even if the exchange of social approval is a primary influence, there are undoubtedly many others as well.

\paragraph{Ethical Considerations.} We recognize that our research could be used by malicious actors to incentivize online hate through positive social reinforcement. We believe the benefits of understanding the social motivators of online hate production outweigh the risks the research poses. This research was IRB-approved.

\section*{Materials and Methods}

\subsection*{Data}
For this research, we use a Twitter dataset of hateful users. Twitter is a platform where users can share tweets and potentially receive engagement and feedback from others, for instance, in the form of likes and replies. Our dataset is based on a dataset collected by \cite{bianchi2022njh}, which collected tweets from 2020-2021 referencing US and UK anti-immigration terms. These tweets are annotated for four sub-types of incivility (profanities, insults, outrage, and character assassination) and two sub-types of intolerance (discrimination and hostility). In their dataset, 18,803 tweets were annotated as uncivil, intolerant, or both. Since only the tweet IDs were provided, we recreated the dataset using our own API credentials in January 2023. We fetched 8,790 tweets (47\% of the original dataset) produced by 7,566 unique users. The rest of the tweets were unavailable for many possible reasons (e.g., a tweet was deleted, the user set their visibility to private, or the account was suspended). Using these 7,566 users who are known to have posted hateful tweets, we collected our dataset of these hateful users’ most recent tweets, up to 3,200 tweets per user (the maximum allowed by Twitter’s historical API). Of the 21 million tweets we collected, 2.9 million of them are original tweets (i.e., not retweets, quotes, or replies), which is the main focus of this research. Note that due to our data collection setup, our dataset contains \textit{only} users who expressed hate at some point during their Twitter tenure, which appropriately limits the scope of this research, yet limits it from generalizing to other users.

For each tweet, our data includes the tweet text and its engagement metrics: like counts, retweet counts, quote counts, and reply counts. Due to data API limitations, we cannot systematically collect the identity of users who engaged with a tweet (e.g., which user liked a post), nor can we collect the original quote or reply messages. We additionally collect each user’s profile description and metadata, including their account age, verification status, follower count, following count, statuses count (the number of total tweets they shared), and listed count (the number of public lists of which the user is a member).

\begin{figure}[t]
    \centering
    \includegraphics[width=0.6\linewidth]{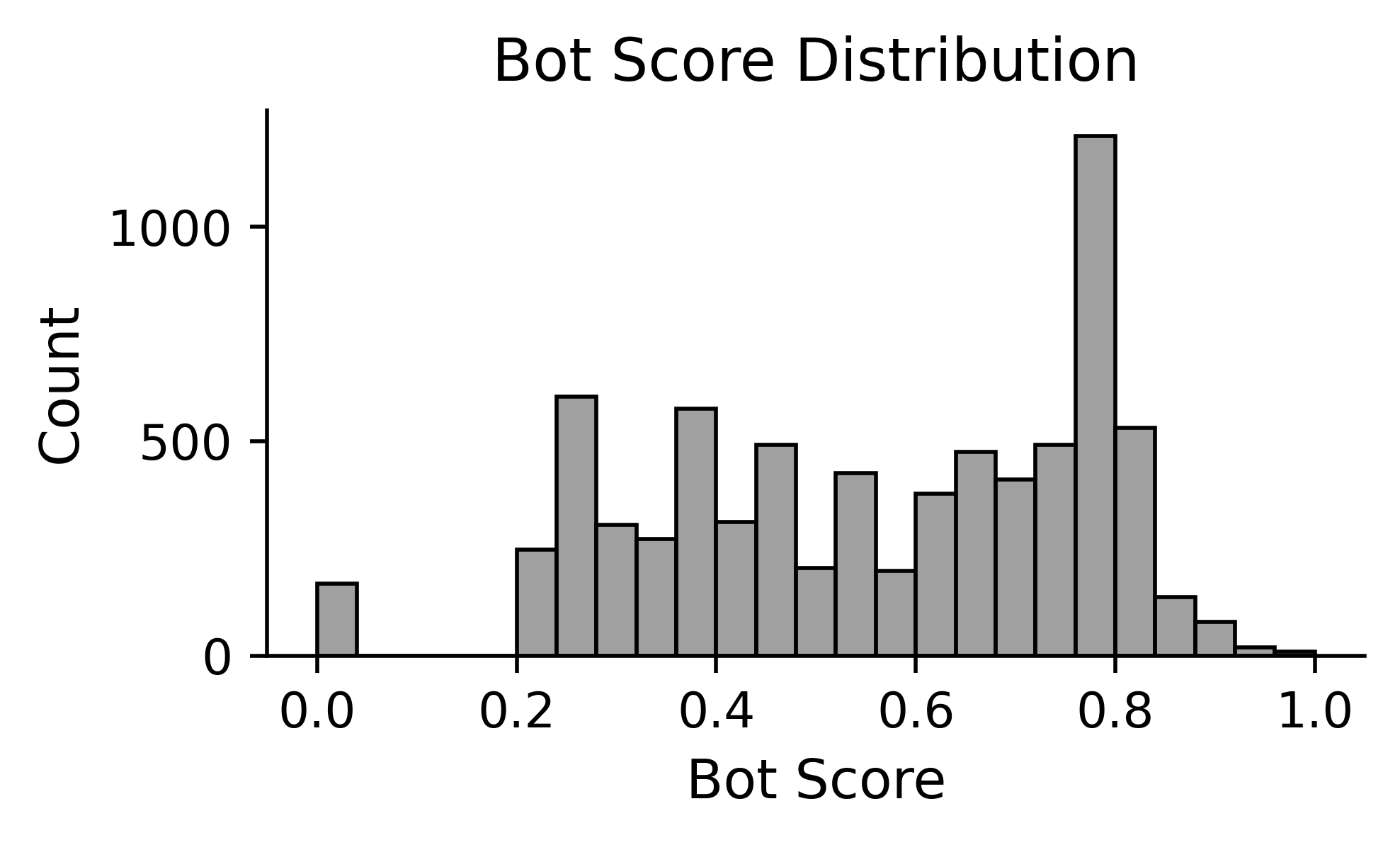}
    % \captionsetup{justification=centering}
    \caption{Distribution of user bot scores.}
    \label{fig:bot_dist}
\end{figure}

\subsection*{Data Preprocessing }

We apply Botometer \cite{yang2022botometer} to detect the presence of inauthentic accounts, scoring each user on a range from 0 to 1, with 1 being most likely to be a bot. Botometer returns two bot scores per user: a language-agnostic score and another that incorporates features derived from the English language. Since the users originally tweeted immigration hate terms in English, we defer to the English-based bot score, though we note that the two scores were very similar. Figure \ref{fig:bot_dist} illustrates the distribution of bot scores per user in this dataset. We can see that this dataset of users likely contains a high proportion of bots, with a mean bot score of 0.56 and a median of 0.58. Therefore, for the rest of this study, we replicate all analyses at two thresholds of bot elimination: users with bot scores $<= 0.8$, which is a conservative choice given the peak in the distribution, and users with bot scores $<= 0.5$. We also remove outlier users, those whose follower counts or engagement metrics exceed three standard deviations from the mean after transformation. We further remove outlier users with transformed follower counts or engagement metric counts that exceed three standard deviations from the mean for each level of bot elimination. The follower count is log-transformed to remove skewness, with 1 added to all counts to avoid 0s when taking the log. The engagement metrics are also log-transformed and further normalized by the log of the follower count to make them comparable across users with varying numbers of followers. Following Frimer et al. \cite{frimer2023incivility}, we also replace all 0s in the engagement metrics with 0.1. See the Supplementary Information for the statistics of these metrics.

After eliminating users with bot scores greater than 0.8 and other statistical outliers, 6,665 users, 2.5 million original tweets, and 19 million total tweets remain. After eliminating users with bot scores greater than 0.5 and other statistical outliers, 2,985 users, 1 million original tweets, and 8.7 million total tweets remain. In both cases, between 1-2\% of users are verified.

Using all 21 million tweets in our dataset, we compile two social networks: a retweet network and a mention network. These networks are compiled using all 21 million tweets. In these networks, users are nodes, and edges represent either a retweet or mention interaction. Mentioning refers to all acts of mentioning (using `@') that are \textit{not} a retweet, which could include quoting, replying, or otherwise referencing the user in a tweet. We disambiguate retweets from mentions because retweeting is usually considered a form of endorsement \cite{boyd2010tweet,metaxas2015retweets} while mentioning could be used to criticize publicly \cite{hemsley2018tweeting}. Each edge also comes with an edge weight, which is equal to the frequency of the retweet or mention between the two users. Details of the networks can be found in Table \ref{tab:toxicity_homophily}.

\begin{figure}[t]
    \centering
    \includegraphics[width=0.7\linewidth]{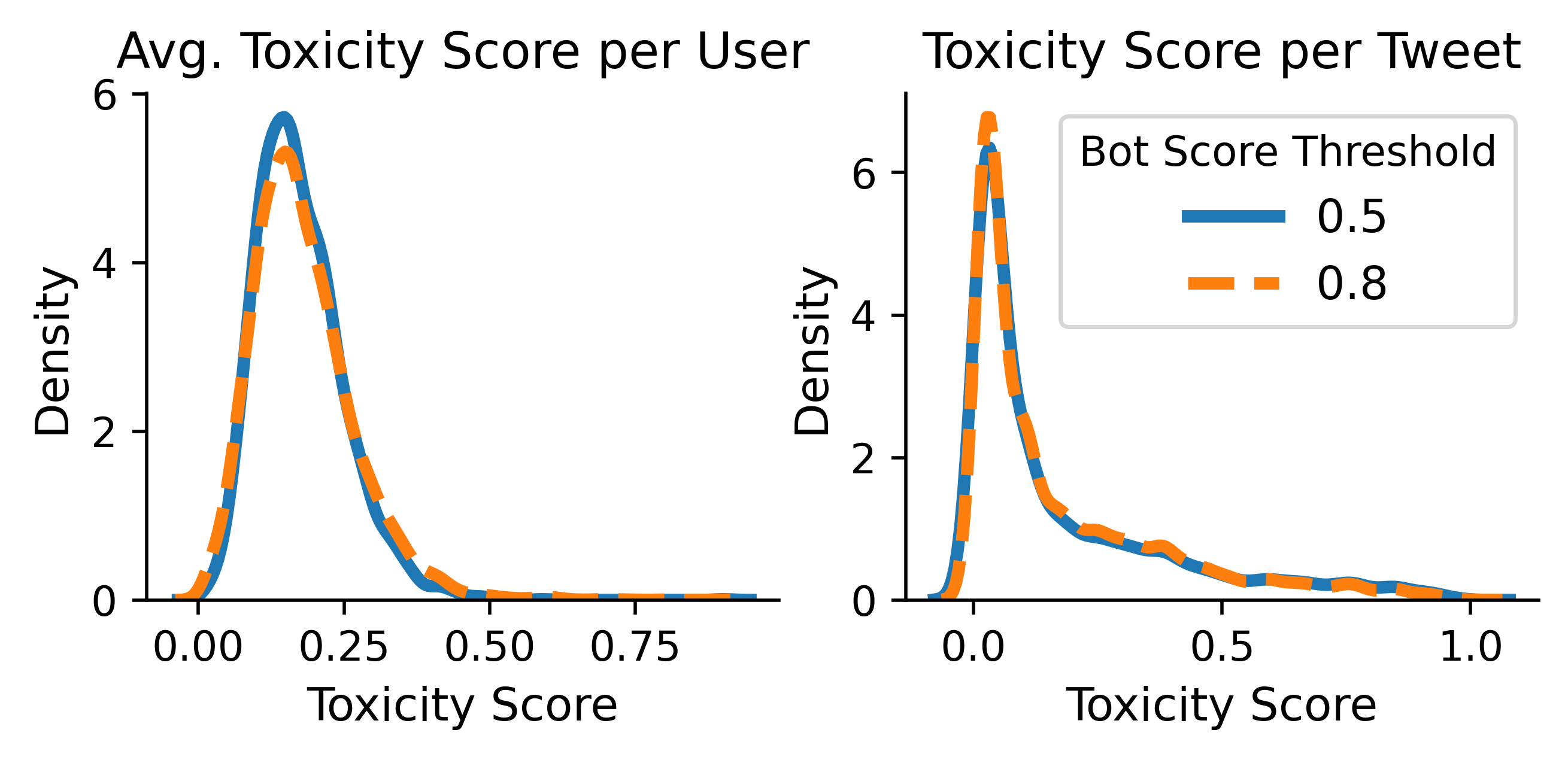}
    \caption{Distribution of the hate scores per user (average hate score of their original tweets) and per original tweet.}
    
    \label{fig:hate_score_dist}
\end{figure}

\subsection*{Measuring Toxicity} 
To detect toxicity, we apply the Perspective API\footnote{\url{https://perspectiveapi.com/}} on every original tweet. The Perspective API is a popular hate detector used in many similar studies \cite{kim2021distorting,frimer2023incivility}. Using its flagship \texttt{Toxicity} score, we compute a hate score per tweet on a scale of 0 (not toxic) to 1 (very toxic). Additionally, we compute a single hate score for each user which is the average of the score scores of all their tweets. Fig. \ref{fig:hate_score_dist} shows the distributions of the tweet-level and user-level score scores.

\subsection*{Measuring Network Homophily}

For the first research objective of determining whether toxic behavior is homophilous, we use the network assortativity method, commonly used to measure how related the edges of a network are in terms of some node attribute \cite{newman2003mixing}, which, in this case, is the average hate score of a user. It is computed as the Pearson correlation of the average hate scores of every pair of users connected by a retweet or mention edges, ranging from $-1$, which indicates that users are preferentially connected to users with the opposite toxicity (disassortative) to $1$, which indicates that users are preferentially connected to other users with similar hate scores (assortative). However, network assortativity does not take into account edge weights, and it neglects to consider that a user’s social network usually consists of multiple connections. Therefore, we also compute the correlation between each user’s hate score and the weighted average toxicity of all of their neighbors using the edge weights in the calculation of that weighted average.

\subsection*{Measuring the Effects of Social Engagement on Toxicity }

The second research objective aims to understand how others’ social engagement with one’s hate message, in the form of likes, retweets, replies, and quotes, affects the hatefulness of one’s subsequent posts. An overly simple solution to this problem could be to binarize the likes, perhaps exploring whether \textit{any} likes or receiving more likes than some $x$ amount would impact toxicity. However, this approach is unsuitable considering that some users in our dataset receive very few likes on their tweets while others are accustomed to receiving hundreds of likes. One potential solution is to scale the social engagement metrics based on users’ follower counts, which can be a reasonable proxy for their popularity. However, we also need to adjust carefully for the varying impact of that social engagement. For instance, an additional ``like'' may carry significantly greater weight for a less popular user than a more popular one.

To address the problems, we categorize tweets into ones receiving ``high'' or ``low'' (or neither) amounts of social engagement by capturing the deviation of actual social engagement from the level of social engagement a user may have expected. We approximate the expected social engagement using the predictive model described below. We then analyze the change in a user's toxicity if the user expected \textit{much less} or \textit{much more} than expected social engagement. 

To facilitate the computational modeling of users, we use several forms of user representations that are based mainly on language features. The first is the \textit{Social-LLM} (Social Large Language Model) user representation model \cite{jiang2023social,retweetbert}, which is a social network representation learning method. The model we use is trained with both retweet and mention edges, as well as users' profile descriptions and metadata features. Social-LLM begins with an LLM representation of the users' profile descriptions and continues to minimize the distance between two users' representations if the users are connected by a retweet or mention interaction. This model, as demonstrated in Jiang et al. \cite{jiang2023social}, can be very useful in detecting the hatefulness of users by combining social network cues with social media content cues. The low-dimensional Social-LLM user embeddings are useful encapsulations of the user's social network features.  In addition, we also use the LLM embedding of the user's profile descriptions (Profile LLM) as well as the LLM of the tweet text (Tweet LLM). Following Jiang et al. \cite{jiang2023social}, we use SBERT-MPNet (\texttt{sentence-transformers/allmpnet-base-v2}) as the LLM for all embeddings.

To calculate the expected social engagement, we train a deep neural network to predict likes, retweets, etc., for every tweet in our dataset based on the following features:

\begin{itemize}
    % best 0.5 model: rt + mn, undirected unweighted, + profile and metadata
    % best 0.8 model: rt + mn weighted, undirected, + profile and metadata
    \item Social-LLM user embeddings \cite{jiang2023social} containing social network and social media content cues.
    \item Profile LLM embedding of the user's profile description 
    \item Tweet LLM embedding
    \item User metadata (follower count, verified, etc.)
    \item Hate score of the tweet
    \item Average hate scores of the past 50 tweets
    \item Average social engagement metrics of the past 50 tweets
    \item Other social engagement metrics this tweet had that are \textit{not} the one being predicted (e.g., using retweets, quotes, and replies to predict likes)
\end{itemize}

% The first set of features, the user embeddings, is particularly important as it is a low-dimensional representation of each user in relation to their social network. The social network is important since, as we will show in \S\ref{sec:rq1}, users' toxic behaviors are heavily linked with their social network. We refer readers to the Appendix for model architecture and evaluation details.

With a model to estimate expected social engagement, we can then calculate the difference between the actual versus expected amount of social engagement. Since we are interested in when a user receives substantially less or more engagement than they may expect, we standardize this difference and look at instances where the z-score of the difference is smaller than $-2$ (less than expected) or more than $2$ (more than expected). We choose $|2|$ because z-scores less than 2 may represent a fluctuation in social engagement that would not be perceived by a user as other than a normal deviation within expectations. In contrast, a higher z-score threshold, such as $|3|$, would result in statistical outliers and also yield too few instances for a meaningful comparison. 

These tweets that experienced a dramatically unexpected amount of social engagement would be referred to as \textit{anchor} tweets. We then compare how the toxicity levels of a user change after the anchor tweet, using varying temporal windows of 30, 50, or 80 tweets before and after the anchor tweet. Here, we operate under the assumption that an unexpected amount of social engagement may alter a user's behavior. As our social engagement data is not timestamped, it is possible that a tweet may have received likes, retweets, etc., after the user has posted the next 30-80 tweets, but we believe this is unlikely. Our assumption is that the majority of social engagement occurred well before users posted an additional 30-80 tweets.

\bibliography{main}

\bibliographystyle{Science}

\section*{Acknowledgements}

This work was supported in part by DARPA under contract number HR001121C0169.
\section*{Supplementary Information}
\subsection*{Data Details}
\begin{table*}[t]
    \centering
    \footnotesize
    
    \caption{The statistics of the number of followers and average numbers of likes, retweets, quotes, and replies per user.}
    \label{tab:metric_stats}
    % \footnotesize
    \begin{tabular}{lrrrrrrrrrrr}
        \toprule
        & & \multicolumn{5}{c}{\textbf{Raw Metrics}} & \multicolumn{5}{c}{\textbf{Transformed}}\\
        \cmidrule(lr){3-7} \cmidrule(lr){8-12}
        & &  Min & Max & Mean & SD & Median   & Min & Max & Mean & SD & Median  \\
        \midrule
        \multicolumn{5}{l}{\textit{Bot Score} $<=0.5$ ($N=2,985$)} \\
        % \midrule
        & Followers &  1 &  1,232,710.00 &  3,359.18 &  32,838.93 &  557.00 &  0.30 &  6.09 &  2.73 &  0.71 &  2.75 \\
        & Likes     &  0 &    13,356.06 &    19.86 &    277.85 &    1.01 & -3.00 &  1.42 & -0.03 &  0.35 &  0.00 \\
        & Retweets  &  0 &     1,327.37 &     2.91 &     30.00 &    0.17 & -3.32 &  1.13 & -0.30 &  0.38 & -0.28 \\
        & Quotes    &  0 &      116.08 &     0.30 &      3.23 &    0.03 & -4.02 &  0.48 & -0.53 &  0.36 & -0.48 \\
        & Replies   &  0 &     1,492.49 &     1.53 &     28.37 &    0.23 & -3.32 &  0.54 & -0.28 &  0.32 & -0.24 \\
        \midrule
        \multicolumn{5}{l}{\textit{Bot Score} $<=0.8$ ($N=6,664)$} \\
        % \midrule
        & Followers &  1 &  3,627,321.00 &  7,093.31 &  86,518.45 &  559.00 &  0.30 &  6.56 &  2.70 &  0.87 &  2.75 \\
        & Likes     &  0 &    37,450.54 &    30.43 &    566.38 &    0.80 & -5.39 &  1.42 & -0.11 &  0.48 & -0.04 \\
        & Retweets  &  0 &     5,320.85 &     5.71 &     88.32 &    0.17 & -7.88 &  1.13 & -0.35 &  0.53 & -0.29 \\
        & Quotes    &  0 &      489.23 &     0.50 &      7.41 &    0.03 & -8.60 &  0.48 & -0.57 &  0.52 & -0.47 \\
        & Replies   &  0 &     1,492.49 &     1.96 &     28.47 &    0.20 & -7.61 &  0.54 & -0.34 &  0.46 & -0.28 \\
        \bottomrule
    \end{tabular}

\end{table*}

\subsubsection*{User Statistics}
Table \ref{tab:metric_stats} displays the raw and transformed social engagement metrics of the users in our dataset.

\begin{figure}
    \centering
    \includegraphics[width=\linewidth]{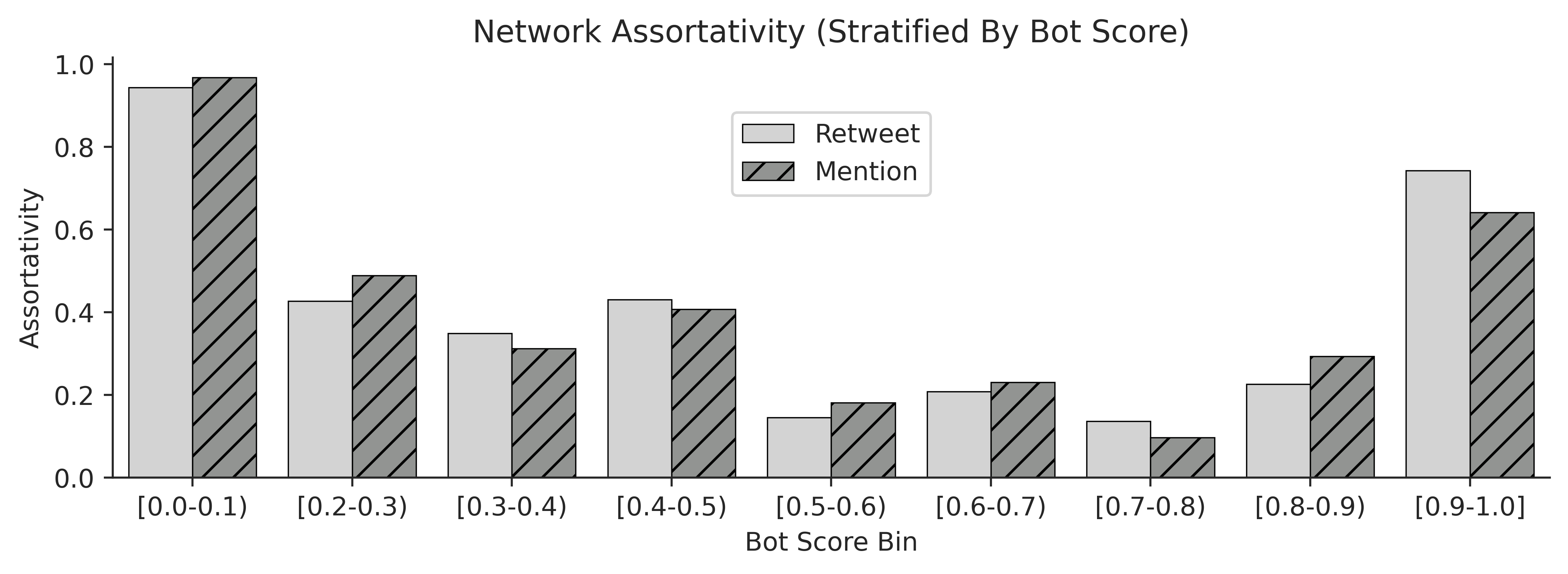}
    \caption{Hate score network assortativity of users in the same bot score bin. All assortativity measures are significant ($p<0.001$), indicating that users are all assortative or homophilous with each other in terms of their hate scores.}
    \label{fig:bot_bin_assortativity}
\end{figure}

\subsubsection*{Bot Score and Homophily in Toxic Behavior}
For completeness, we analyze the relationship between a user's bot tendencies and their homophily in-depth. We divide the users by their bot scores in 0.1 increments and measure the homophily in toxic behavior among users in the same bot score bin using network assortativity in Fig. \ref{fig:bot_bin_assortativity}. Note that there are no users with bot scores between 0.1 and 0.2. We calculate the network assortativity of both the retweet and mention networks, which are all significant ($p<0.0001$). We find that hate score network assortativity is high for both the very human-like users (bot score bin $[0.0, 0.1)$) and the very bot-like  (bot score bin $[0.9, 1.0]$) and low otherwise. One explanation is that bots and humans do not preferentially attach themselves to each other, but rather, bots interact mostly with bots and humans interact mostly with humans (see \cite{williams2020homophily}). This finding motivates our choice to separate analyses in this work by bot scores.

\subsection*{Social Engagement Experiment Details}

\subsubsection*{Social Engagement Estimation Model}
We select the best Social-LLM models from RQ1 for use in the RQ2 modeling of social engagement estimation. The best models based on test performance all incorporated retweet edges, mention edges, profile description, and user metadata but differed regarding how the network feature was used. For users with bot threshold $=0.5$, the best model used the network edges as undirected and unweighted. For users with bot threshold $=0.8$, the best model used network edges as undirected but weighted. We use the following hyperparameters/settings when training the social engagement prediction model: learning rate $=0.001$, number of epochs $=20$, number of dense layers $=4$, hidden units $=768$, batch size $=32$, ReLU activation, and batch normalization.

% best 0.5 model: rt + mn, undirected unweighted, + profile and metadata
% best 0.8 model: rt + mn weighted, undirected, + profile and metadata

For users with bot threshold $=0.5$, our trained models produce $R^2$ values of 0.577, 0.515, 0.371, and 0.273 for predicting the number of likes, retweets, replies, and quotes, respectively, per tweet. For users with bot threshold $=0.8$, our trained models produce $R^2$ values of 0.531, 0.380, 0.167, and 0.467 for predicting the number of likes, retweets, replies, and quotes, respectively, per tweet. 

\begin{table*}
    \centering
    \footnotesize
    
    \caption{The change in hate score when a user received lower vs. higher than expected amount of likes, retweets, replies, or quotes over a window of $k=50$ tweets. We test the significance of the difference between the distributions using a Mann-Whitney U test ($**p<0.01$).}
    \label{tab:res_bot08}
    \begin{tabular}{lllllllllccc}
        \toprule
          & \multicolumn{3}{c}{\textbf{Lower Than Predicted}} & \multicolumn{3}{c}{\textbf{Higher Than Predicted}}& \\
         \cmidrule(lr){2-4} \cmidrule(lr){5-7}
         Metric & \# Ex & \# Users & $\Delta$ Hate & \# Ex & \# Users & $\Delta$ Hate &  Diff. in $\Delta $ Hate & Sig.  \\
         \midrule
         Likes & 2,397 & 748 & -0.001421 & 82,488 & 1,900 & 0.000204 & 0.001624  & ** &  \\
         Retweets & 67 & 52 & 0.006862 & 143,592 & 2,537 & 0.000073 & -0.006789 & N.S&  \\
        Replies & 4 & 1 & -0.004567 & 105,176 & 2,746 & -0.000169 & 0.004398 & N.S. & \\
        Quote & 14,693 & 1,034 & 0.000789 & 157,379 & 3,656 & 0.000148 & -0.000641& ** & \\
         \midrule
    \end{tabular}
    
\end{table*}
\subsection*{Additional Results}
\subsubsection*{Results with Higher Bot Threshold}
We report the results for users with bot scores $<=0.8$ in Table \ref{tab:res_bot08} using a time window of $k=50$. We note that due to the change in data distribution, there is no longer a sufficient number of samples where users experienced significantly fewer retweets and replies. Therefore, we cannot adequately analyze how retweets and replies affect the toxic behaviors of these users. The results for likes and quotes corroborate our main findings. When users experience more likes than expected instead of fewer likes than expected, they will become more toxic in the future. However, if they receive more quotes than expected instead of fewer quotes than expected, they will become less toxic in the future. Likes act as a positive social reinforcement, whereas quotes act as a negative social reinforcement. 

\begin{figure}[t]
    \centering
    \includegraphics[width=0.7\linewidth]{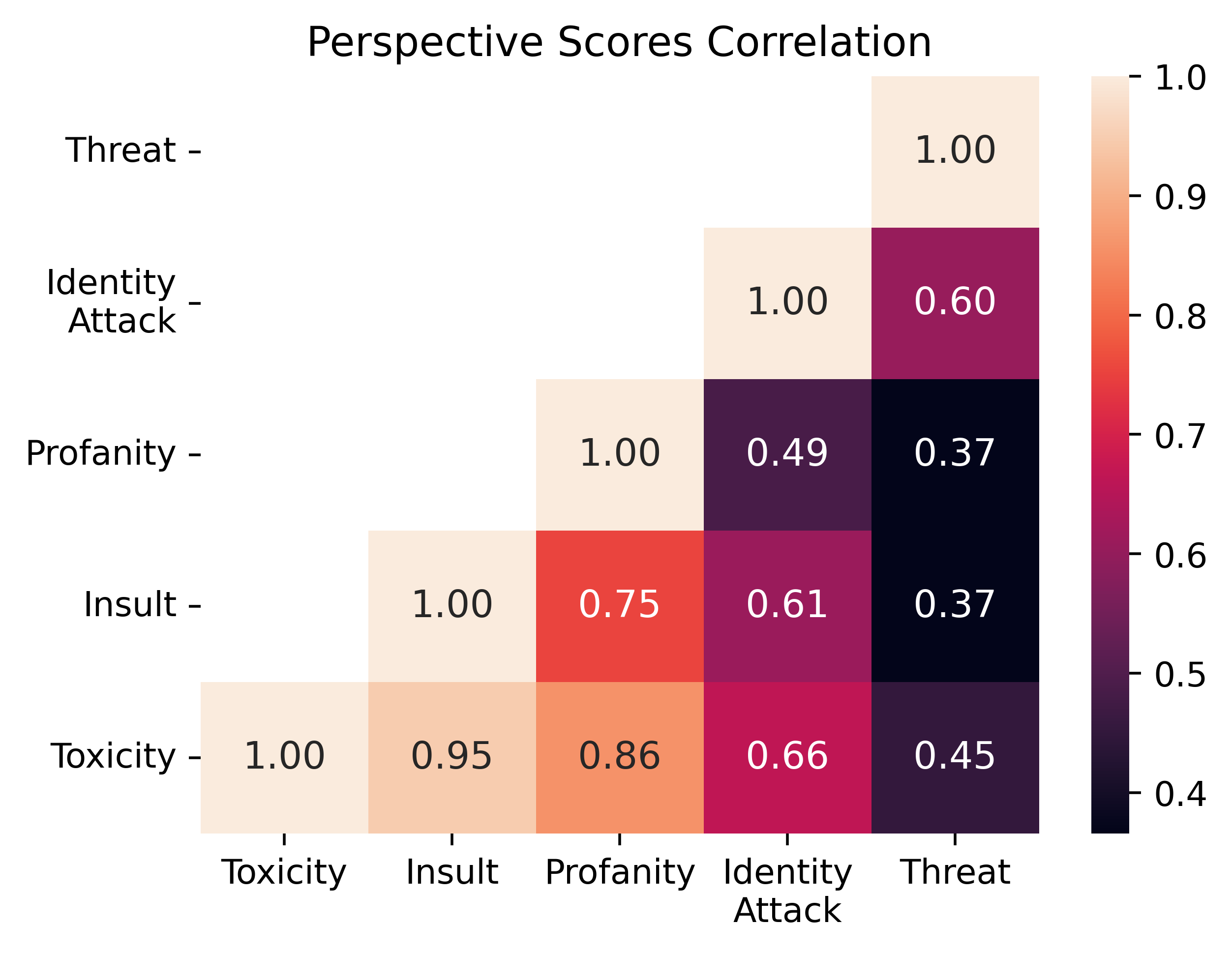}
    \caption{Pearson correlation between two Perspective scores of each tweet. All values are statistically significant ($p<0.05$).}
    \label{fig:perspective_corr}
\end{figure}

\begin{figure}[t]
    \centering
    \includegraphics[width=0.7\linewidth]{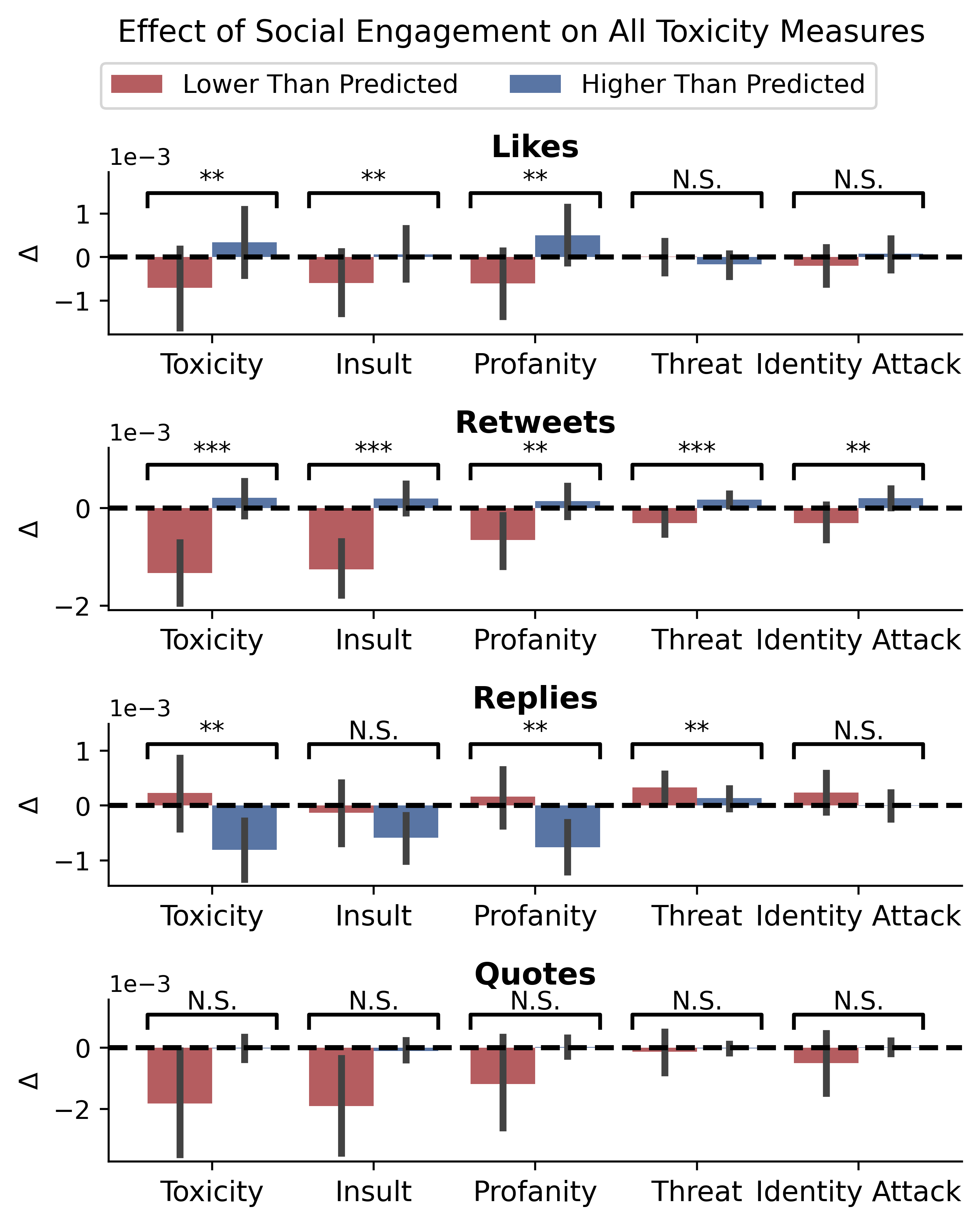}
    \caption{We display the effect of social engagement on all five toxicity attributes at $k=50$, similar to Fig. \ref{fig:social_engagement_toxicity_diff}.}
    \label{fig:all_toxicity}
\end{figure}

\subsubsection*{Robustness Checks with Other Toxicity Measures} 
Besides the flagship \texttt{TOXICITY} score, the Perspective API also computes \texttt{IDENTITY\_ATTACK}, \texttt{INSULT}, \texttt{PROFANITY}, and \texttt{THREAT} scores, which are additional hateful messaging measures we use in robustness checks. We display the correlations among all five toxicity attributes of the tweets in our dataset in Fig \ref{fig:perspective_corr}. Most similar to \texttt{TOXICITY} is \texttt{INSULT} at $r=0.95$ ($p<0.001$), and the least similar is \texttt{THREAT} at a $r=0.45$ ($p<0.001$).

To ensure the validity of our results, we repeat our analysis with the four other toxicity attributes in Fig. \ref{fig:all_toxicity} as a robustness check. The positive impact of likes on increased toxicity also occurs in \texttt{INSULT} and \texttt{PROFANITY}. The reduction of toxicity due to more replies is also found in \texttt{PROFANITY} and \texttt{THREAT}. Interestingly, retweets positively affect subsequent toxicity in all five toxicity attributes, further corroborating our contention about the potency of retweets.
\end{document}